\documentclass[aps,twocolumn,prd,nofootinbib]{revtex4}

\usepackage{amsmath}
\usepackage{graphicx}
\usepackage{dcolumn}
\usepackage{bm}
\usepackage{amssymb}
\usepackage{latexsym}

\begin{document}

\title{Phantom inflation with a
steplike potential}

\author{Zhi-Guo Liu\footnote{Email: liuzhiguo08@mails.gucas.ac.cn}}
\author{Jun Zhang\footnote{Email: junzhang34@gmail.com}}
\author{Yun-Song Piao\footnote{Email: yspiao@gucas.ac.cn}}

\affiliation{College of Physical Sciences, Graduate University of
Chinese Academy of Sciences, Beijing 100049, China}

\begin{abstract}

The phantom inflation predicts a slightly blue spectrum of tensor
perturbation, which might be tested in coming observations. In
normal inflation models, the introduction of step in its potential
generally results in an oscillation in the primordial power
spectrum of curvature perturbation. We will check whether there is
the similar case in the phantom inflation with steplike potential.
We find that for same potentials, the oscillation of the spectrum
of phantom inflation is nearly same with that of normal inflation,
the difference between them is the tilt of power spectrum.

\end{abstract}

\maketitle

%\hskip 1.6cm PACS number(s): {98.80.Cq} \vskip 0.4cm

\section{Introduction}

The results of recent observations are consistent with an adiabatic
and nearly scale invariant spectrum of primordial perturbations, as
predicted by the simplest models of inflation. The inflation is
supposed to have taken place at the earlier moments of the universe
\cite{G},\cite{LAS},\cite{S1}, which superluminally stretched a tiny
patch to become our observable universe today. During the inflation
the quantum fluctuations in the horizon will be able to leave the
horizon and become the primordial perturbations responsible for the
structure formation of observable universe. In this sense, exploring
different inflation models is still an interesting issue.

Recently, the phantom field, for which the parameter of state
equation $\omega<-1$ and the weak energy condition is violated, has
acquired increasely attention \cite{RC}, inspired by the wide use of
such fields to describe dark energy, e.g.
\cite{CKW},\cite{HL},\cite{NOV},\cite{vx},
\cite{ST},\cite{GPZ},\cite{ACL}. The simplest realization of phantom
field is a normal scalar field with reverse sign in its dynamical
term. The quantum theory of such a field may suffer from the
causality and stability problems \cite{CHT,CJM}. However, this does
not mean that the phantom field is unacceptable. Actions with
phantomlike form may be arise in supergravity \cite{N}, scalar
tensor gravity \cite{BEPS}, higher derivative gravity \cite{P},
braneworld
 \cite{SS}, stringy \cite{FM}, and
other scenarios \cite{CL,S}. The phantom inflation, which is
drived by the phantom field, has been proposed \cite{YZ}, and
widely studied in
\cite{phan},\cite{GJ},\cite{NO},\cite{BFM},\cite{WY},\cite{Y},\cite{Fe}.
In phantom inflation, the power spectrum of curvature perturbation
can be nearly scale invariant. The duality of primordial spectrum
to that of normal field cosmology has been studied in \cite{phan,
Piao2004}. In the meantime the spectrum of tensor perturbation is
slightly blue tilt, which is distinguished from that of normal
inflationary models \cite{YZ}.

%The exit mechanisms have been studied in \cite{PR,G,PZ,GD,ST,Fe}.

The power spectrum of perturbations in the normal inflation depends
on the inflaton potential. In some inflationary models, there may be
inflaton potentials with a large number of steps. The steplike
change in the potential will result in a universal oscillation in
the spectrum of primordial perturbations
\cite{Starobinsky},\cite{Adams:1997de},\cite{step},\cite{Leach},\cite{ACE},\cite{Hamann:2007pa},\cite{JSS},\cite{LM},
and also \cite{PFZ},\cite{Cai}.
%showed that supergravity-inspired models may
%give rise to inflaton potentials with a large number of steps.
%Each step corresponds to a symmetry breaking phase transition in a
%field coupled to the inflaton. The inflaton mass then changes
%suddenly when each transition occurs. These steps can create
%oscillating features in the primordial power spectrum
%Oscillation will also be directly imprinted on the inflaton
%potential itself.
A burst of oscillations in the primordial spectrum seems to provide
a better fit to the CMB angular power spectrum
\cite{J3S},\cite{HAJ}. In this Letter, we will check whether there
is a similar feature in the phantom inflation with steplike
potential. We will consider a quadratic potential with a step and a
hybrid potential with a step, respectively, and numerically
calculate the corresponding power spectrum, and then compare them
with that of normal inflationary model.

%we consider a class of inflationary models which is driven by
%phantom scalar field and a quadratic potential with a step. The
%inflation is not interrupted but the effect on the density
%perturbation is very important.

The plan of this work is as follows. In section II we present the
phantom inflation scenario and give the simple analysis of the
background evolution with steplike potential. In section III we
discuss the calculation of the power spectrum and present numerical
results for the primordial spectrum. Section IV contains discussion
and conclusions. Note that we work in units such that $\hbar =c=8\pi
G=1$.

\section{The Background of phantom inflation with a step }

%The special character of phantom inflation requires the inflation
%paradigm to be suitably redesigned.

The simplest realization of phantom field is a normal scalar field
with reverse sign in its dynamical term. This reverse sign results
in that, different from the evolution of normal scalar field
during the normal inflation, the phantom field during the phantom
inflation will be driven by its potential up along its potential,
e.g.\cite{ST, GPZ}. Thus if initially the phantom field is in the
bottom of potential, analogous to the slow rolling regime of the
normal scalar field, the phantom field will upclimbe and enter
slow climbing regime. Hereafter, the phantom inflation begins, and
after some time the phantom inflation ends and the universe enters
into a period dominated by the radiation. The exit from the
phantom inflation can be implemented by introducing an additional
normal scalar field \cite{YZ}, in which the exit mechanism is
similar to the case of hybrid inflation \cite{linde94, cope94},
the imposition of backreaction \cite{WY}, the wormhole \cite{GJ},
the brane/flux annihilation in string theory \cite{Y}, or the
nonminimally coupling of the phantom to gravity \cite{Fe}.

%Consider a spatially flat Friedmann-Robertson-Walker(FRW) spacetime
%with the unperturbed metric
%\begin{eqnarray}
%ds^2=-dt^2+a(t)^2\delta_{ij}dx^idx^j
%\end{eqnarray}
%The effective action of simple phantom field as follows
%\begin{eqnarray}
%S=\int
%d^4x\sqrt{-g}[\frac{1}{2}(\partial_{\mu}\varphi)^2-V(\varphi)]
%\end{eqnarray}
%We consider a spatially homogenous but time-dependent background
%field $\phi$, the energy density $\rho$ and pressure $p$ can be
%expressed as
%\begin{eqnarray}\label{p}
%\rho=-\frac{1}{2}\dot{\phi}^2+V(\phi)
%  ~~,~~ p=-\frac{1}{2}\dot{\phi}^2-V(\phi)
%\end{eqnarray}
%From (\ref{p}), we can see that the state parameter $\omega\equiv
%\frac{p}{\rho} <-1$ for $\rho >0$.
When the phantom field is minimally coupled to the gravitational
field, the Friedmann equation can be written as
\begin{eqnarray}\label{H}
3H^2=-\frac{1}{2}\dot{\phi}^2+V(\phi),
\end{eqnarray}
where $H^2$ is positive, which means in all cases for the phantom
evolution its dynamical energy must be smaller than its potential
energy, thus the phantom inflation is not generally interrupted by
the step in despite of the height of step. This can be compared with
that in normal inflation, in which it is possible that
$\dot{\phi}^2>V(\phi)$ for a high step and the inflation is
interrupted for a short interval. The phantom field satisfies the
equation
%
%\begin{eqnarray}\label{phi}
$\ddot{\phi}+3H\dot{\phi}-V'(\phi)=0$.
%\end{eqnarray}
%
The phantom field is driven to upclimbe along its potential is
reflected in the minus before $V^\prime$ term. Define the
slow-climb parameters \cite{YZ}
\begin{eqnarray}
\epsilon_{pha}\equiv -\frac{\dot H}{H^2}~~~,~~~
\delta_{pha}\equiv-\frac{\ddot\phi}{H\dot\phi}
\end{eqnarray}
when the conditions $|\epsilon_{pha}|<<1$ and $|\delta_{pha}|<<1$
are satisfied, Eq (\ref{H}) can be solved semianalytically. Then we
have $a\sim e^{Ht}$ approximately, which means the universe enters
into the inflationary phase driven by the phantom field.

The numerical solutions of evolution equations are required for
accurately evaluating the perturbation spectrum. We shift the
independent variable to $\alpha=\ln{a}$, which will facilitate the
numerical integration. With this replacement and using the energy
conservation equation, we have
\begin{eqnarray}\label{HH}
H_\alpha=\frac{1}{2}H\phi_\alpha^2,
\end{eqnarray}
\begin{eqnarray}\label{scalar}
\phi_{\alpha\alpha}+(\frac{H_\alpha}{H}+3)\phi_\alpha-\frac{1}{H^2}V'=0.
\end{eqnarray}
where the subscript $\alpha$ denotes differentiation for $\alpha$
and the prime denotes differentiation with the scalar field
$\phi$.

In general, the step in the potential can be modelled by
introducing the term proportional to
$\tanh(\frac{\phi-\phi_{step}}{\delta})$. We consider a simple
inflaton potential $m^2\phi^2$, then this potential with a step
can be given by
\begin{eqnarray}\label{potential}
V(\phi)=\frac{1}{2}m^2\phi^2\left(1+\beta\tanh(\frac{\phi-\phi_{step}}{\delta})\right).
\end{eqnarray}
This potential has a step at $\phi=\phi_{step}$ with size and
gradient governed by $\beta$ and $\delta$. We will focus on small
features in the potential, and thus will limit the parameter
$\beta$ small. In this case, the phantom will upclimbe
continuously through the step while the inflation will not be
ceased.

\begin{figure}[htbp]
\includegraphics[scale=0.6,width=8.5cm]{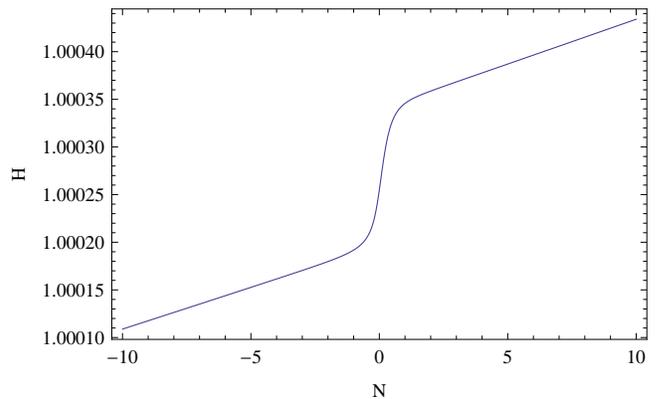}
\caption{Evolution of the Hubble parameter $H$. $H$ is in the unit
of $H0$ which is a model dependent parameter. $H$ increases and is
different from normal inflation } \label{fig:h}
\end{figure}

The numerical results with the potential (\ref{potential}) can be
seen in Figure \ref{fig:h}. We can see that $H$ is nearly constant
in the inflationary era, but there are some differences from the
normal background. $H$ increases slowly in the phantom inflation
compared with decrease slowly in the normal inflation, since in the
phantom inflation the energy density is increased. There is a very
abrupt change due to the existence of the step.

\begin{figure}[htbp]
\includegraphics[scale=0.6,width=8.5cm]{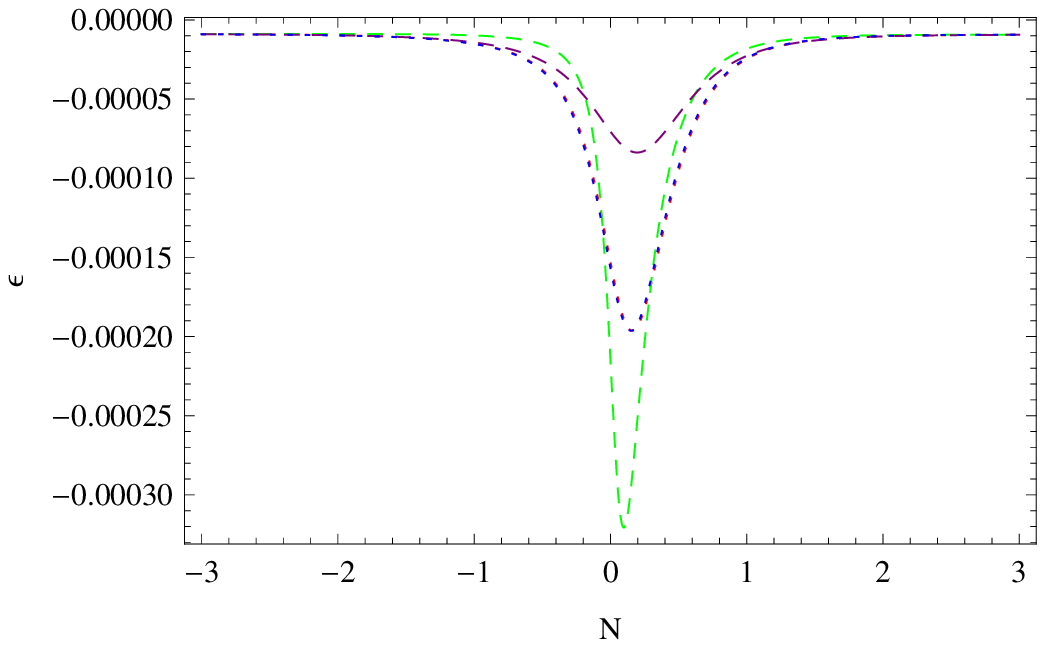}
\includegraphics[scale=0.6,width=8.5cm]{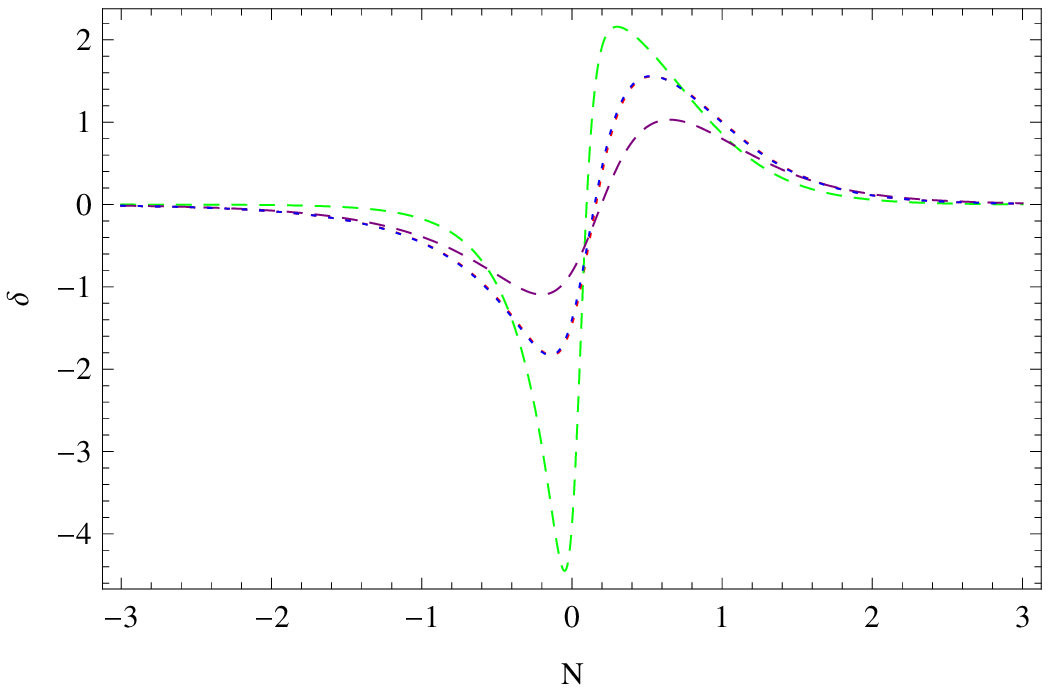}
\caption{Evolution of the two slow-climb parameters $\epsilon_{pha}$
(top) and $\delta_{pha}$ (bottom) with the introduction of the step.
Different of the $\beta$ and $\delta$ parameters: $\beta=10^{-2},
\delta=10^{-3}$ (red dotted line), $\beta=10^{-2},
\delta=5\times10^{-4}$ (green dashed line), $\beta=5\times10^{-3},
\delta=10^{-3}$ (purple dashed line), $\beta=5\times10^{-3},
\delta=5\times10^{-4}$ (blue dotted line).} \label{fig:epde}
\end{figure}

The slow-climb parameters in term of $\alpha$ can be changed as
follows
\begin{eqnarray}
\epsilon_{pha}=-\frac{H_\alpha}{H}~~,~~
\delta_{pha}=-\frac{\phi_{\alpha\alpha}}{\phi_\alpha}-\frac{H_\alpha}{H}.
\end{eqnarray}
The introduction of the step leads to a deviation from slow-climb
inflation, though during this interval it is still inflation. We
have plotted the evolution of the two slow-climb parameters
$\epsilon_{pha}$ and $\delta_{pha}$ around the time when the field
crosses the step in Figure \ref{fig:epde}.

\section{The perturbation of phantom inflation with a step}

We will investigate the power spectrum of curvature perturbations
during the phantom inflation. In the scalar case it is
advantageous to define a gauge invariant potential
\cite{M},\cite{KS}
\begin{equation}
u = -z \mathcal{R}
\end{equation}
where $z\equiv a\sqrt{2|\epsilon_{pha}|}$, and $a$ denotes the scale
factor, $H$ is the Hubble parameter, and the dot is the derivative
with respect to the physical time $t$.

The equation of motion $u_{\rm k}$ in the momentum space is
\begin{equation}  \label{uevln}
u_{\rm k}'' + \left( k^2-\frac{z''}{z} \right)u_{\rm k}=0,
\end{equation}
where the prime denotes differentiation with respect to conformal
time and $k$ is the wave number. The solution depends on the
relative sizes of $k^2$ and $z''/z$. The $z''/z$ term can be
expressed as $2a^2H^2$ plus terms that are small during the
phantom inflation. In general, on subhorizon scales $k^2 \gg
z''/z$, the solution of perturbation is a plane wave
\begin{equation}  \label{freesc}
u_{{\rm k}} \rightarrow \frac{1}{\sqrt{2k}} \, e^{-ik\tau}\,.
\end{equation}
The $\frac{1}{\sqrt{2k}}$ is obtained by the quantization of mode
function $u_k$. Here a normal quantization condition has actually
been applied, like that in normal field, which seems inconsequential
for phantom field. However, it is generally thought that the phantom
field might be only the approximative simulation of a fundamental
theory below certain physical cutoff, and the full theory should be
well quantized. In another viewpoint, we might assume that initially
there is not phantom field, thus the perturbation deep inside the
horizon follows normal quantization condition, then the evolution
with $w<1$ emerges for a period, which is simulated
phenomenologically by the phantom field, as in island cosmology
\cite{island} or \cite{Y}. Thus the primordial perturbation induced
by the phantom fields has to have a normal quantization condition as
its initial condition, or it cannot be matched to that of initial
background. While on superhorizon scales $k^2 \ll z''/z$ the
dominated mode is
\begin{equation}
u_{{\rm k}} \propto z\ \label{eqn:regmode}
\end{equation}
which means that the curvature perturbation
\begin{equation}
|{\mathcal R}_{{\rm k}}|=|u_{{\rm k}}/z|,
\end{equation}
is constant.

The spectrum ${\cal P}_{{\cal R}}(k)$ is defined as
\begin{equation}
\langle {\cal R}_{{\rm k}_1} {\cal R}^*_{{\rm k}_2} \rangle =
        \frac{2\pi^2}{k^3} {\cal P}_{{\cal R}} \delta^{3} \,
        (k_1- k_2) \,,
\end{equation}
and is given by
\begin{equation}
\label{pspec} {\cal P}_{{\cal R}}^{1/2}(k) =
\sqrt{\frac{k^3}{2\pi^2}} \,
        \left| \frac{u_{{\rm k}}}{z} \right| \,.
\end{equation}
We will numerically calculate ${\cal P}_{{\cal R}}$. However,
before this it is interesting to show the result of ${\cal
P}_{{\cal R}}$ in the slow climbing approximation, which is
\begin{eqnarray}
\mathcal {P}_\mathcal
{R}=\frac{1}{2|\epsilon_{pha}|}\left(\frac{H}{2\pi}\right)^2
\left({k\over aH}\right)^{n_\mathcal {R}-1}, \label{p}\end{eqnarray}
where the spectrum index is $n_\mathcal {R}-1\simeq
-4\epsilon_{pha}+2\delta_{pha}$. This power spectrum may be either
blue or red. The results are dependent on the relative magnitude of
$\epsilon_{pha}$ and $\delta_{pha}$, see \cite{YZ} for the details.
Eq.(\ref{p}) is valid only when the slow climb approximation is
satisfied, i.e.$|\epsilon_{pha}|\ll 1$ and $|\delta_{pha}|\ll 1$,
however, when the potential has a sharp step, the deviation of
$\delta$ is large, see Fig \ref{fig:epde}. In this case, we have to
evolve the full mode equation numerically without any
approximations.

The perturbation equation (\ref{uevln}), with the replacement
$\alpha=lna$, can be written as
\begin{eqnarray}
u_{\alpha\alpha}+(1+\frac{H_\alpha}{H})u_\alpha+\left(\frac{k^2}{e^{2\alpha}H^2}-\frac{z''/z}{e^{2\alpha}H^2}\right)u=0
\end{eqnarray}
with
\begin{eqnarray}
\frac{z''}{z}=a^2H^2\left(2-5\frac{H_\alpha}{H}-2\frac{H_\alpha^2}{H^2}-
4\frac{H_\alpha}{H}\frac{\phi_{\alpha\alpha}}{\phi_\alpha}+\frac{V''}{H^2}\right)
\end{eqnarray}
The evolution of spectrum is governed by the competition between
the $k^2$ and $z''/z$ terms. The overall normalization of
$\mathcal {P}_\mathcal {R}$ is proportional to $m^2$,
$\phi_{step}$ determines the wavelength at which the feature
appears. The dominant contribution to $z''/z$ is from the $V''$
term and is proportional to $\beta/\delta^2$. Thus the range of
$k$ affected by the feature roughly depends on the square root of
$\beta/\delta^2$.

\begin{figure}[htbp]
\includegraphics[scale=0.6,width=8.5cm]{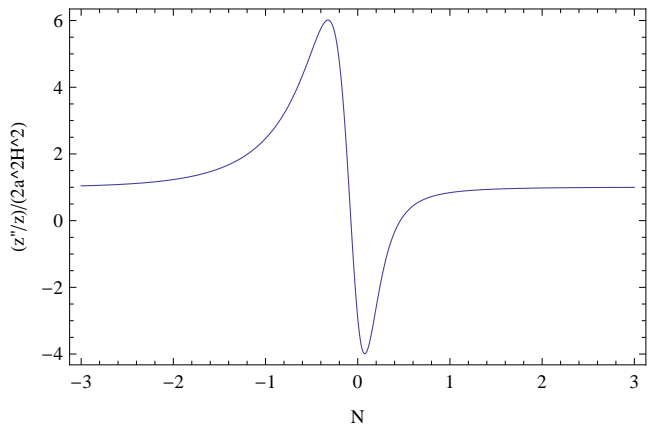}
\caption{Evolution of $z''/z$ for $\beta=0.01$ and $\delta=0.001$
with the efolding number of inflation N and we have set N=0 at the
step in the potential} \label{fig:z}
\end{figure}
\begin{figure}[htbp]
\includegraphics[scale=0.6,width=8.5cm]{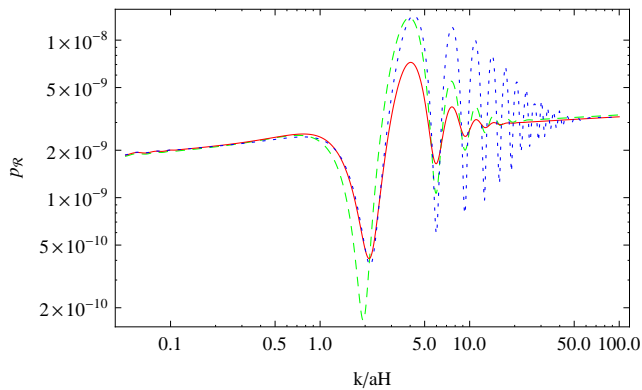}
\caption{The power spectrum of curvature perturbation for the
phantom inflation with the potential (\ref{potential}),
$\beta=5\times10^{-3},\delta=4\times10^{-2}$(true red
line),$\beta=5\times10^{-3},\delta=10^{-2}$(dotted blue
line),$\beta=10^{-2},\delta=4\times10^{-2}$(dashed green line).}
\label{fig:pmp}
\end{figure}
We plot the $z''/z$ term in Fig \ref{fig:z} and solve the equation
numerically and give the results of power spectrum in Fig
\ref{fig:pmp}. In Fig \ref{fig:z}, we can see that the $z''/z$ term
is very different from $2a^2H^2$. It has a sharp oscillation near
the step. Fig \ref{fig:pmp} shows the power spectrum of phantom
inflation with potential of Eq.(\ref{potential}). We plot it with
three group parameters and can see that the introduction of step
results in a large deviation from slow climbing inflation and then a
burst of oscillations superimposed on the nearly scale invariant
scalar power spectrum. The magnitude and extent of oscillation is
dependent on the amplitude and gradient of the step. Thus as in
normal inflation, these oscillations will inevitably leave
interesting imprints in the CMB angular power spectrum, which might
provide a better fit e.g.\cite{J3S},\cite{HAJ}.

\begin{figure}[htbp]
\includegraphics[scale=0.6,width=8.5cm]{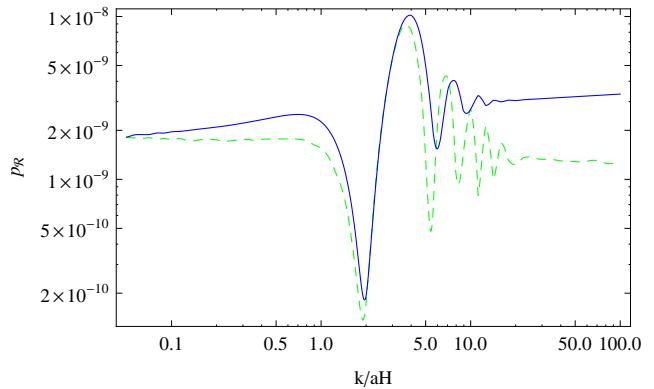}
\caption{Effects of a step in the potential on the power spectrum of
curvature perturbation for the phantom inflation and the normal
inflation, respectively. The blue solid line corresponds to the
phantom inflation with $m=7.5\times10^{-6},
\beta=10^{-2},\delta=5\times10^{-2}$ and the green dashed line
corresponds to the normal inflation with
$m=7.5\times10^{-6},\beta=10^{-2},\delta=2\times10^{-2}$ }
\label{fig:pn}
\end{figure}

\begin{figure}[htbp]
\includegraphics[scale=0.6,width=8.5cm]{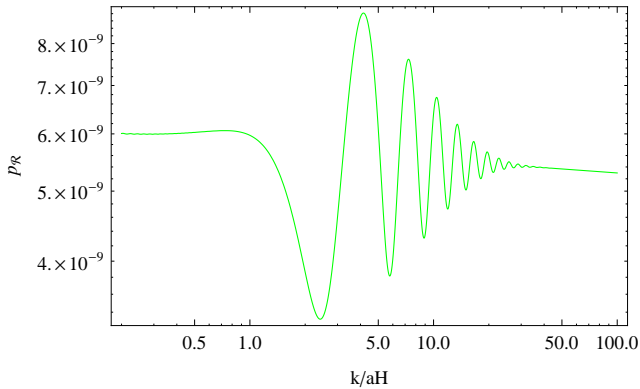}
\caption{The power spectrum of curvature perturbation for the
phantom inflation with the potential (\ref{hybrid}),
$V_0=3.7\times10^{-14}, m=3.2\times10^{-8}, \phi_{step}=0.0125,
\beta=5\times10^{-4},\delta=10^{-5}$. Corresponding to the spectra
index $n_s<1$.} \label{fig:v0p}
\end{figure}

We compare the primordial spectrum of phantom inflation and normal
inflation in Fig \ref{fig:pn}, the top line shows the primordial
power spectrum of phantom inflation for the potential with
$m=7.5\times10^{-6},\beta=10^{-2},\delta=5\times10^{-2}$. The normal
inflation model with a same potential has a red spectrum
$n_{\mathcal {R}}<1$, however, the phantom inflation background
model has a blue spectrum $n_{\mathcal {R}}>1$. This is because what
we consider here is the potential with $m^2\phi^2$. In principle, we
can have the models of phantom inflation with $n_{\mathcal {R}}<1$
by choosing a suitable potential \cite{YZ}. We can consider a model
of phantom inflation with the potential
\begin{eqnarray}\label{hybrid}
V(\phi)=V_0+\frac{1}{2}m^2\phi^2\left(1+\beta
\tanh(\frac{\phi-\phi_{step}}{\delta})\right)
\end{eqnarray}
where $V_0$ is constant which dominates the potential. This
potential is same with that of hybrid inflation with normal field.
Fig \ref{fig:v0p} shows that $n_{\mathcal {R}}<1$ which is similar
to that of the normal inflation with $m^2\phi^2$.

The power spectrum of tensor perturbation is only dependent on
$\epsilon_{pha}$. We set the parameter $\beta$ small, thus actually
$|\epsilon_{pha}|\ll 1$ around the step. In this case, the tensor
spectrum is hardly affected, which remain nearly scale invariant.
However, due to $\epsilon_{pha}<0$, thus $n_T\simeq
-2\epsilon_{pha}$ is slightly blue tilt for the phantom inflation
\cite{YZ}, which is distinguished from the normal inflation, in
which $n_T$ is generally red tilt.

\section{Conclusion and Discussion}

The phantom field can naturally appear in effective actions of
some theories, which might be only the approximative simulation of
a fundamental theory below certain physical cutoff. Thus the
phantom cosmology have been widely studied. The phantom inflation
predicts a slightly blue spectrum of tensor perturbation, which is
distinguished from that of the normal inflation, in which the
tensor perturbation is generally red tilt. This is a smokegun for
the phantom inflation, which might be tested in coming
observations.

In normal inflation models, the introduction of step in its
potential generally results in an oscillation in the primordial
power spectrum of curvature perturbation. In this Letter, we find
that for same potentials with the step, the oscillation of the
spectrum of phantom inflation can be nearly same with that of normal
inflation, and the magnitude and extent of oscillation is dependent
on the amplitude and gradient of the step. The difference between
them is the tilt of power spectrum. However, the same tilt can be
obtained by considering a different potential of the phantom
inflation.

In general, ${\dot\phi}^2$ for the phantom must be smaller than
its potential energy in all time. Thus the phantom inflation is
not interrupted by the step in despite of the height of step. This
can be compared with that in normal inflation, in which it is
possible that for a high step the inflation is interrupted for a
short interval. Here, we have limited the parameter $\beta$ small,
however, it is interesting to consider the phenomena of $\beta\gg
1$, i.e. there is a large step, by which the density of dark
energy observed might be linked to that of inflation, as in the
eternal expanding cyclic scenario, e.g.\cite{FLPZ},\cite{IBF}.
This might lead to a lower CMB quadrupole in observable universe
if the step is just in the position of potential, in which the
perturbation with Hubble scale leaves the horizon during the
phantom inflation, as in the bounce inflation model
\cite{PFZ},\cite{Cai}. It is possible that the phantom inflation
with steplike potential can be effectively implemented in certain
warped compactifications with the brane/flux annihilation,
e.g.\cite{Y}. We expect to back to the relevant issues in the
coming works.

\textbf{Acknowledgments}

This work is supported in part by NSFC under Grant No:10775180,
11075205, in part by the Scientific Research Fund of
GUCAS(NO:055101BM03), in part by National Basic Research Program
of China, No:2010CB832804

%We thank xxxxxxxxxx. This work is supported xxxxxxx

\end{document}